\documentclass[a4paper]{jpconf}
\usepackage{graphicx}
\usepackage{svg}
\usepackage{amsmath}
\usepackage{subcaption}
\usepackage{floatrow}
\usepackage[normalem]{ulem}

\begin{document}
\title{Confirming the high pressure phase diagram of the Shastry-Sutherland model}

\author[1]{Yuqing Ge $^{1,*}$, Daniel Andreica $^2$, Yasmine Sassa $^1$, Elisabetta Nocerino $^3$, Ekaterina Pomjakushina $^4$, Rustem Khasanov $^5$, Henrik M. Rønnow $^6$, Martin Månsson $^3$, and Ola Kenji Forslund $^{1,\dag}$}

\address{$^1$Department of Physics, Chalmers University of Technology, Göteborg 41296, Sweden}
\address{$^2$Faculty of Physics, Babes-Bolyai University, 400084 Cluj-Napoca, Romania}
\address{$^3$Department of Applied Physics, KTH Royal Institute of Technology, SE-106 91 Stockholm, Sweden}
\address{$^4$Laboratory for Multiscale Materials Experiments, Paul Scherrer Institut, PSI, Villigen, Switzerland}
\address{$^5$Laboratory for Muon Spin Spectroscopy, Paul Scherrer Institut, Villigen, Switzerland}
\address{$^6$Lab. for Quantum Magnetism (LQM), École Polytechnique Fédérale de Lausanne, Switzerland}

\ead{yuqing.ge@chalmers.se *,  ola.forslund@chalmers.se $\dag$}

\begin{abstract}
A Muon Spin Rotation ($\mu^+SR$) study was conducted to investigate the magnetic properties of SrCu$_2$(BO$_3$)$_2$ (SCBO) as a function of temperature/pressure. Measurements in zero field and transverse field confirm the absence of long range magnetic order at high pressures and low temperatures. These measurements suggest changes in the Cu spin fluctuations characteristics above 21 kbar, consistent with the formation of a plaquette phase as previously suggested by inelastic neutron scattering measurements. SCBO is the only known realisation of the Shatry-Sutherland model, thus the ground state mediating the dimer and antiferromagnetic phase is likekly to be a plaquette state.

\end{abstract}

\section{Introduction}
Low dimensional magnetism is a field of considerable interest as it offers opportunities to study new and exciting physics. The study of 2-dimensional (2D) magnetism has developed rapidly during the recent years. Compounds having a layered crystal structures with weak inter-layer interaction enable the possibility of studying 2D magnetism in bulk materials \cite{cortie2020two}. 
One famous model for a 2D magnetic system is the Shastry-Sutherland (SS) model, which forsees an orthogonal dimer network of spin $S = 1/2$ \cite{shastry1981exact} with exchange interactions between the nearest-neighbours ($J$) and the next-nearest-neighbours ($J'$). Based on this frustrated spin system, a variety of ground states is predicted depending on the ratio $\alpha = J'/J$. For $J'/J < 0.5$, the system is in a dimer phase where the nearest-neighbouring spins form spin-singlets. On the other side for $J'/J>1$, a 2D antiferromagnetic (AFM) phase with significant quantum fluctuations is expected \cite{dalla2015fractional}. However, the ground state in the intermediate region ($0.5 < J'/J < 1$) has been debated and various possibilities have been suggested theoretically: AFM order \cite{muller2000exact,shastry1981exact}, helical order \cite{albrecht1996first} or even columnar dimer \cite{zheng2001phase}. 

\begin{figure}[ht]
\includegraphics[width = 34pc]{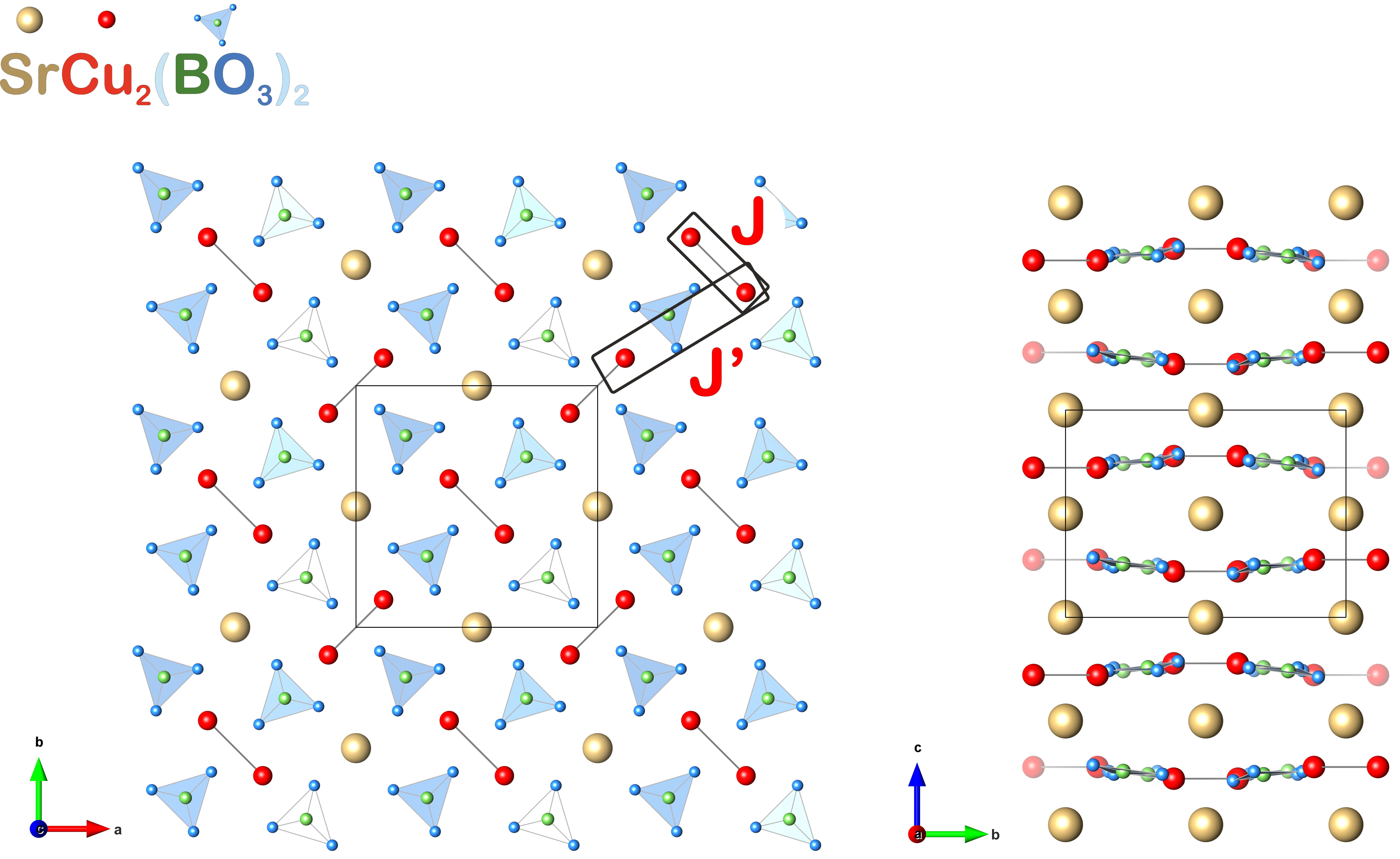}
\caption{Cross section in ab-plane(left panel) and bc-plane(right panel). Left panel shows the intra-dimer interaction $J$, and the inter-dimer interaction $J'$.  }
\label{fig:Intro}
\end{figure}

SrCu$_2$(BO$_3$)$_2$ (SCBO) is the only known realization of the SS model, as its 2D crystal structure (figure \ref{fig:Intro}) lends itself well to the kind of spin system described by the SS. The structure consists of alternating layers of Sr atoms and a Cu$_2$(BO$_3$)$_2$ layer of Cu/triangular BO$_3$ system. The physics in this compound is determined by the interactions between the Cu ions. They are arranged in an orthogonal rectangular network where competing interactions are established between nearest neighbour and next-nearest neighbour ions. The Cu ions are dimerised below 8 K at ambient pressure, and the intra-dimer interaction is mediated by a super-exchange interaction through the O atoms \cite{radtke2015magnetic}.
In order to investigate the intermediate state of the SS model, it is necessary to be able to tune the ratio $\alpha = J'/J$. Here, hydrostatic pressure presents a clean way of controlling $\alpha$ \cite{radtke2015magnetic}. Therefore, a series of hydrostatic pressure studies on SCBO were initiated, including neutron scattering \cite{zayed20174,haravifard2014emergence}, specific heat capacity \cite{jimenez2021quantum}, tunnel diode oscillator \cite{shi2022discovery} and electron spin
resonance \cite{sakurai2010development} experiments. Among these experiments, a recent inelastic neutron scattering \cite{zayed20174} determined the phase above 21.5~kbar to be a plaquette phase. Then a heat capacity study \cite{jimenez2021quantum} shows clear sign of a critical point around 20 kbar, and an Ising transition into the plaquette phase at 1.8 K.

In this work, we studied the material by high-pressure $\mu ^+$SR up to 24 kbar. The absence of long range magnetic order up to this pressure is confirmed by measurements in zero field (ZF). Moreover, measurements in transverse field (TF) confirms the changes in the dynamical character above 21 kbar. This is consistent with the formation of plaquette state, in which the spin correlation is different from that of a dimerised spin system.

\section{Experimental details}
The experiment was performed at the Swiss Muon Source ($S\mu S$), Paul Scherrer Institut, using the General Purpose Decay-Channel (GPD) Spectrometer \cite{khasanov2022perspective}. 
The powder sample of SCBO was filled in a piston-cylinder pressure cell made of MP35N and pressurized. 
Since the sample has a response to muons comparable to that of the pressure cell, it is difficult to distinguish the two contributions when analysing the time spectrum.
Thus, the muons were implanted at two different momenta (100 MeV/c, and 107 MeV/c) where 100 MeV/c resulted in muons stopping only at the pressure cell, while with 107 MeV/c the muons probed also the sample. 
Through this measurement protocol, the sample and pressure cell contributions could be separated in a global fit procedure. Temperature scans from 0.25 K up to 10 K was performed under ZF and TF=50~G at pressures of 0, 21, and 24 kbar. The $\mu^+SR$ data was analysed using the software musrfit \cite{suter2012musrfit}.

\section{Results and discussions}

\begin{figure}
    \centering
    \includegraphics[width=36pc]{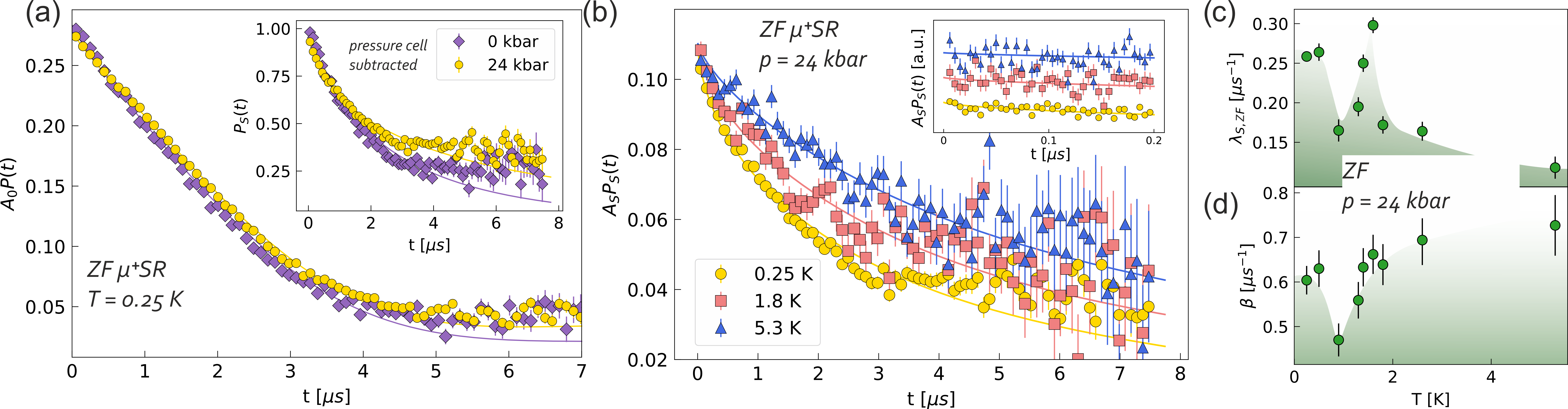}
    \caption{\label{fig:ZF} (a) ZF time spectra collected at $T=0.25$~K for pressures 0, and 24~kbar. Solid curves represent best fits using Eq.~(\ref{eq:ZF}). The inset shows pressure cell subtracted polarization function. The pressure cell parameters were assumed to be pressure independent, and the same configurations were used in analyzing 0 and 24 kbar spectra. (b) Pressure cell subtracted ZF muon time spectra collected at $p=24$~kbar for $T= 0.25$, 0.9, 1.6, and 5.3~K. Solid curves represent best fits using Eq.~(\ref{eq:ZF}). (c) Relaxation rate ($\lambda_{S,ZF}$) as a function of temperature at $p=24$~kbar. (d) Stretched exponent ($\beta$) as a function of temperature for $p=24$~kbar. The green shadow underneath the experimental points is a guide to the eye.}
\end{figure}

The collected ZF $\mu^+$SR time spectra at $T = 0.25$~K for 0 and 24~kbar are shown in Fig.~\ref{fig:ZF}a. Small but notable difference is observed in the initial part of the spectra. The absence of oscillations confirms that no long range magnetic order is formed. Instead, Gaussian Kubo-Toyabe (GKT) and exponential like polarisation is observed. Therefore, the time spectra were fitted using:

\begin{equation}
    \begin{split}
        A_0\mathrm{P_{ZF}}(t) = A_S \mathrm{exp}(-(\lambda_{S,ZF}t)^\beta) + A_{PC}\mathrm{exp}(-\lambda _{PC}t)[\frac{1}{3}+\frac{2}{3}(1-\sigma^2t^2)\mathrm{exp}(-\sigma^2t^2/2)]
    \end{split}
    \label{eq:ZF}
\end{equation}
where $A_0$ is the total asymmetry, and $P_{ZF}(t)$ is the muon depolarisation function under ZF. $A_S$ is the asymmetry, $\lambda_{S,ZF}$ is the relaxation rate and $\beta$ is the power of the stretched exponential for the sample, while  $A_{PC}$ is the asymmetry, $\sigma$ is the static depolarization rate under ZF, and $\lambda_{PC}$ the dynamic (electronic) relaxation rate of the pressure cell \cite{khasanov2022perspective}.

The pressure cell subtracted ZF time spectra for selected temperatures are shown in Fig.~\ref{fig:ZF}(b). As the system is cooled down from $5.3$~K, the spectra changes from a weak to a strongly relaxing exponential.  The temperature dependencies of the ZF fit parameters obtained using Eq.~(\ref{eq:ZF}) of the sample are shown in Fig.~\ref{fig:ZF}(c,d). The relaxation rate exhibits low values at high temperature but slowly increases as the transition temperature $T_C=1.8$~K is approached. The peak observed in $\lambda_{S,ZF}$ suggest the system is in a phase transition to the plaquette states at $T_C$. This is consistent with the transition temperature reported from heat capacity measurements \cite{jimenez2021quantum}.

The stretched exponent ($\beta$) clearly decreases from around 0.8 at $T = 5.3$~K to 0.5 at $T = 0.9$~K (Fig.~\ref{fig:ZF}(d)). A value of $\beta = 1 $ corresponds to a simple exponential function, which could be interpreted as a single Lorentzian field distribution or a dynamic system with other forms of distributions such as a Gaussian, and a value smaller than 1 implies a broader distribution of relaxation rates \cite{yaouanc2011muon}. It is common for $\beta$ to decrease from 1 at high temperature to around $1/3$ at low temperature close to phase transitions \cite{campbell1994dynamics, ogielski1985dynamics, keren1996probing, Forslund2020}. Here, below $T_C$, $\beta$ gradually decreases towards 0.5. Below 0.7 K however, $\beta$ experience a sudden increase and deviates from 0.5. Most likely, this is a spurious effect due to the difficulty in fitting an exponential like relaxation response in an MP35N cell, particularly at low temperatures because of the pressure cell background \cite{khasanov2022perspective}. Another scenario includes the formation of a different spin state of the system at these temperatures. However, such kind of transition is not observed in heat capacity measurements \cite{jimenez2021quantum}.



The sample was also measured under TF = 50~G. TF spectra were fitted with a single exponentially relaxing cosine function:

\begin{equation}
    \begin{split}
    A_0\mathrm{P_{TF}(t)} = &A_{TF}\mathrm{exp}(-\lambda _{TF}t)\mathrm{cos}(2\pi f_{TF}t+\phi_{TF})
    \end{split}
    \label{eq:TF}
\end{equation}
where $A_0$ is the total asymmetry, and $P_{TF}(t)$ is the muon depolarisation function under TF. Both the sample and the pressure cell contributions are modelled together in a single exponentially relaxing cosine function: $A_{TF}$ is the asymmetry, $\lambda_{TF}$ the relaxation rate, $f_{TF}$ the precession frequency under applied field, and $\phi_{TF}$ the phase.

\begin{figure}[ht]
    \centering
    \includegraphics[width=28pc]{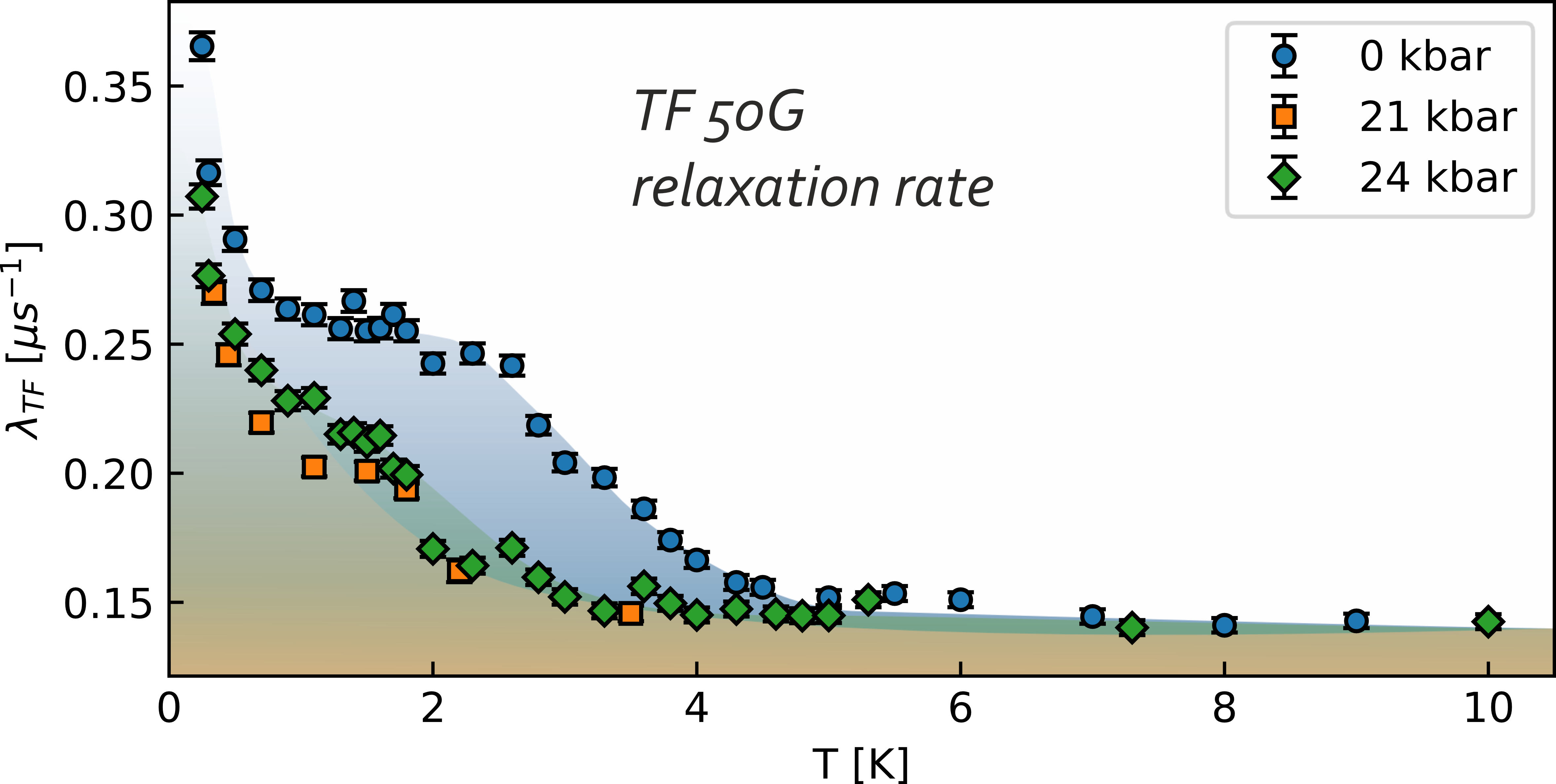}
    \caption{TF relaxation rate ($\lambda_{S,TF}$) as a function of temperature at $p=0$, 21 and 24~kbar, obtained from Eq.~(\ref{eq:TF}). Shadow as a guide for the eye. }
    \label{fig:lambda_TF}
\end{figure}

The obtained TF relaxation rate as a function of temperature, collected at $p=0,\ 21$ and $24$ kbar, is shown in Fig.~\ref{fig:lambda_TF}. At high temperatures, the relaxation rates show no pressure or temperature dependence and is settled at a value of $\lambda_{\rm TF}=0.15~\mu$s$^{-1}$. Below 4~K, $\lambda_{\rm TF}$ starts to increase with decreasing temperature for all pressures. The dependency on temperature at $0$ kbar is consistent with previous $\mu^+SR$ experiment at ambient pressure \cite{sassa2018mu+}. 
However, the rate of increase at $21$ and $24$ kbar is different from the one at $0$ kbar. This pressure dependence is attributed to changes in the local field dynamics and/or the field distribution widths in the sample. Since $\lambda_TF$ is essentially the integrated field correlation function, this change is consistent with the formation of a new spin configuration with its own spin-spin correlation function. Therefore, the qualitative difference in the dynamical nature above 20 kbar is consistent with the appearance of an extra magnon mode above this pressure seen with inelastic neutron scattering \cite{zayed20174}, which was interpreted to be due to the formation of a plaquette state. Finally, the sharp increase below 0.7~K in $\lambda_{\rm TF}$ is a contribution from the pressure cell \cite{GPD}.

\section{Summary}
In summary, we used high pressure $\mu^+ SR$ to confirm the pressure/temperature phase diagram of SCBO. The absence of oscillation in the ZF spectra under higher pressure confirms that no long range magnetic order is formed. Moreover, measurements in TF suggest that the dynamical character changes above 21 kbar. These results are consistent with the formation of a plaquette state above 21 kbar, as suggested in previous studies.

\section{Acknowledgements}

This experiments was conducted in the GPD muon spectrometer, at the Swiss Muon Source (SµS) of the Paul Scherrer Institut (PSI), Villigen, Switzerland, and we are thankful to the staff for their support. The proposal number of the regarded experiment is 20131703. We are grateful to Christian Rüegg for valuable discussions. All fitting of the $\mu^+SR$ data was perform with musrfit \cite{suter2012musrfit}. All images involving crystal structure were made with the VESTA software \cite{momma2011vesta}.
This research was supported by the European Commission through a Marie Skłodowska-Curie Action and the Swedish Research Council - VR (Dnr. 2014-6426 and 2016-06955) as well as the Carl Tryggers Foundation for Scientific Research (CTS-18:272).  Y.S. and O.K.F are funded by the Swedish Research Council (VR) through a Starting Grant  (Dnr. 2017-05078). E.N. and Y.G. are supported by the Swedish Foundation for Strategic Research (SSF) within the Swedish national graduate school in neutron scattering (SwedNess). Y.S. acknowledges funding from the Area of Advance-Material Sciences from Chalmers University of Technology. D.A. acknowledges partial financial support from the Romanian UEFISCDI Project No. PN-III-P4-ID-PCCF-2016-0112. H. M. R. acknowledges funding from Swiss National Science Foundation of project grant number 188648.

\section*{References}

\bibliographystyle{iopart-num}
\bibliography{ref}

\end{document}